\pdfoutput=1
\documentclass[letterpaper, 10 pt, conference]{ieeeconf}  

\IEEEoverridecommandlockouts                              

\let\labelindent\relax
\usepackage[inline]{enumitem}

\usepackage{graphicx}
\usepackage{graphics} 
\usepackage{mathptmx} 
\usepackage{times} 
\usepackage{amsmath} 
\usepackage{amssymb}  

\usepackage{graphicx}
\usepackage{caption}
\usepackage{subcaption}
\usepackage[linesnumbered,ruled,vlined,algo2e]{algorithm2e}
\SetKwRepeat{Do}{do}{while} 
\usepackage{dirtytalk} 
\usepackage{fancybox} 
\usepackage{cite}
\usepackage{balance}
\usepackage{hyperref}
\usepackage{commath}
\usepackage{stmaryrd}

\usepackage{url}

\usepackage{color} 
\usepackage{tikz}
\usetikzlibrary{positioning}

\title{\LARGE \bf
Learning Dynamic-Objective Policies from a Class of Optimal Trajectories
}
\author{
    Christopher Iliffe Sprague\authorrefmark{1}\thanks{\authorrefmark{1} Robotics, Perception and Learning Lab., School of Electrical Engineering and Computer Science, Royal Institute of Technology (KTH), SE-100 44 Stockholm, Sweden} \\ \small\texttt{sprague@kth.se} \and 
    Dario Izzo\authorrefmark{2}\thanks{\authorrefmark{2} Advanced Concepts Team, European Space Technology Center (ESTEC), Noordwijk, The Netherlands} \\ \small\texttt{dario.izzo@esa.int} \and 
    Petter \"Ogren\authorrefmark{1}  \\ \small\texttt{petter@kth.se}
}
\begin{document}
\maketitle
\thispagestyle{empty}
\pagestyle{empty}

\begin{abstract}
Optimal state-feedback controllers, capable of changing between different objective functions,
are advantageous to systems in which unexpected situations may arise.
However, synthesising such controllers, even for a single objective, is a demanding process.
In this paper, we present a novel and straightforward approach to synthesising these policies through a combination of trajectory optimisation, homotopy continuation, and imitation learning.
We use numerical continuation to efficiently generate optimal demonstrations across several objectives and boundary conditions, and use these to train our policies.
Additionally, we demonstrate the ability of our policies to effectively learn families of optimal state-feedback controllers, which can be used to change objective functions online.
We illustrate this approach across two trajectory optimisation problems, an inverted pendulum swingup and a spacecraft orbit transfer, and show that the synthesised policies, when evaluated in simulation, produce trajectories that are near-optimal.
These results indicate the benefit of trajectory optimisation and homotopy continuation to the synthesis of controllers in dynamic-objective contexts.



\end{abstract}


\tikzset{%
  every neuron/.style={
    circle,
    draw,
    minimum size=0.3cm
  },
  neuron missing/.style={
    draw=none, 
    scale=0.9,
    text height=0.27cm,
    execute at begin node=\color{black}$\vdots$
  },
}

\section{Introduction}
An optimal controller that is capable of changing its objective function in the middle of a mission can be very useful in an autonomous system. Imagine a vehicle that started off on a time-optimal trajectory towards a given destination but suddenly finds itself having less energy than expected; perhaps due to a battery failure, a broken solar panel, or a leaking fuel tank. It can then instantly switch to an effort-optimal controller. Conversely, a vehicle that was saving energy for future transitions might suddenly be in a hurry to reach a destination and can then instantly switch to a controller that reaches the goal in the shortest possible time.

In general, applying an optimal control policy is desirable, but often infeasible for many autonomous systems.
The reason for this is that, for many systems characterised by nonlinear dynamics, finding a closed-loop solution to the underlying optimal control is intractable.
A common approach to circumvent this is to precompute an open-loop solution, through trajectory optimisation \cite{betts2010practical}, and apply a simple state-feedback controller to track it.
However, such an approach suffers when unforeseen events require a change of the objective function or lead the system's state out of the controller's region of attraction.

An alternative approach is to apply model predictive control (MPC), where open-loop solutions are incrementally computed and executed on a finite-time horizon. 
Although this approach provides reactivity in response to unforeseen events, it might suffer in performance, as it does not consider the full-time horizon; and in computational complexity, as it requires repeatedly solving an optimisation problem.

\begin{figure}[t]

    \centering
    \begin{subfigure}[b]{0.49\linewidth}
       \includegraphics[width=1\linewidth]{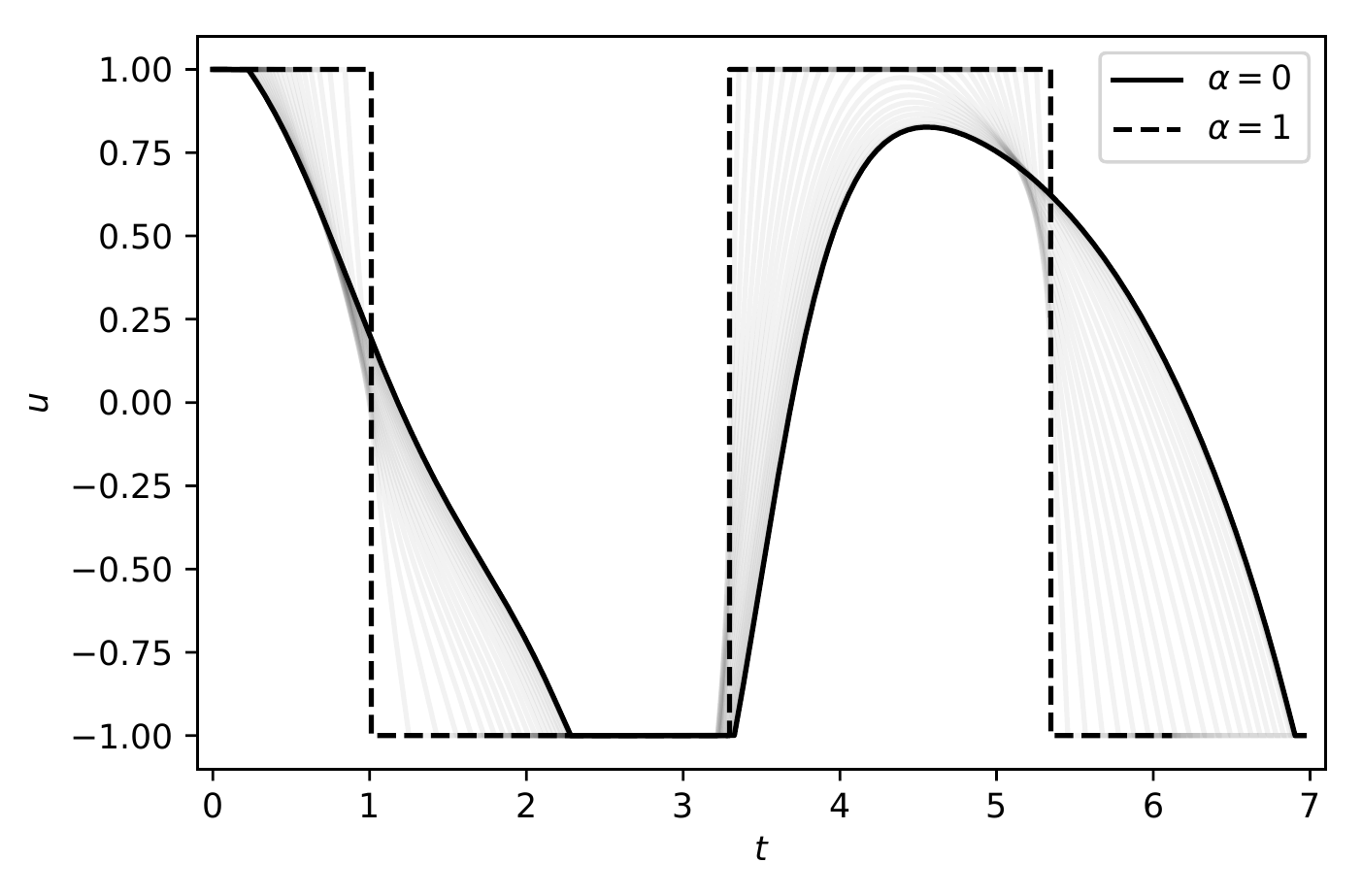}
       \caption{}
        \label{fig:pendulum_control_homotopy}
    \end{subfigure}
    \begin{subfigure}[b]{0.49\linewidth}
       \includegraphics[width=1\linewidth]{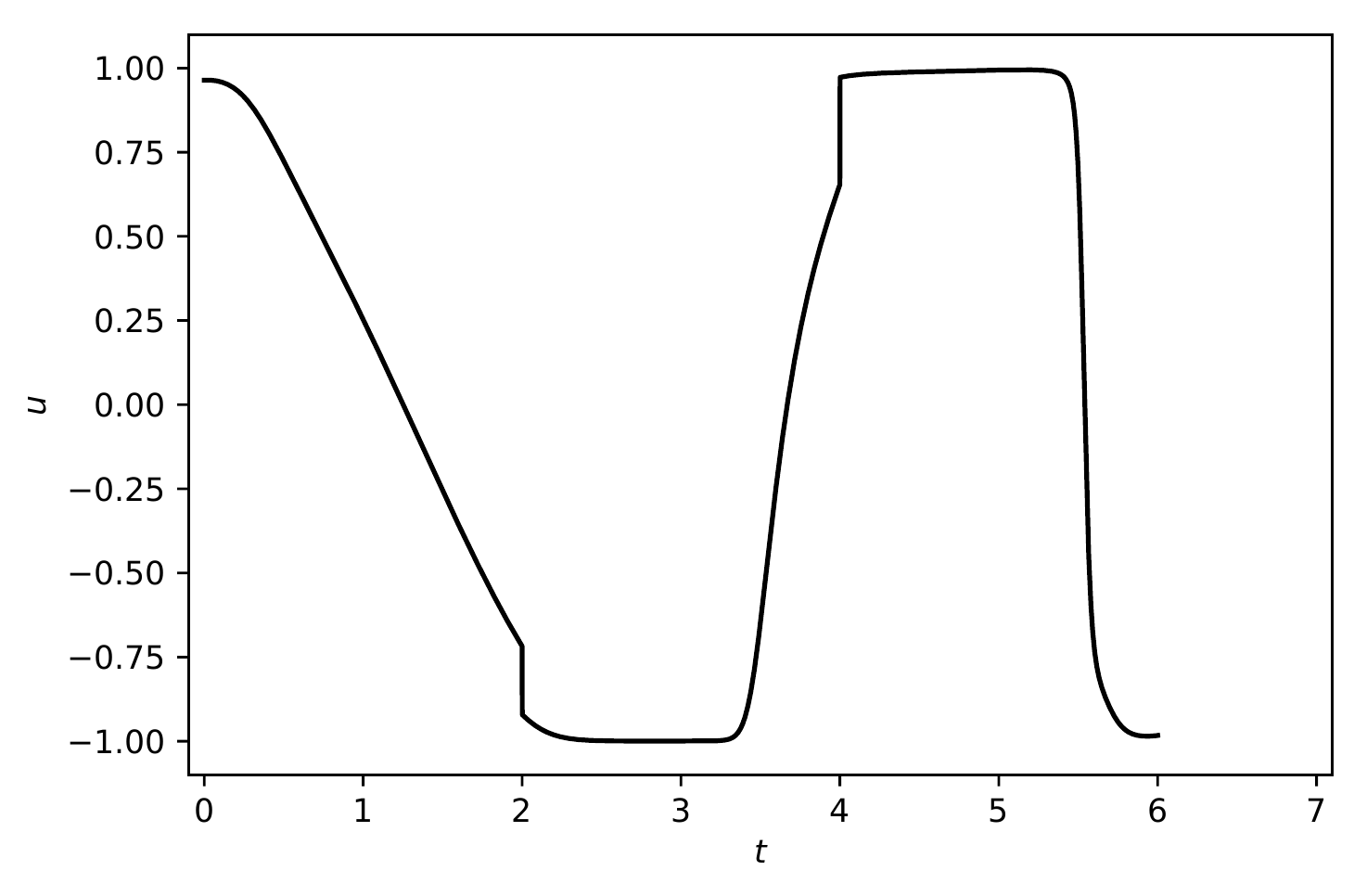}
       \caption{}
    \end{subfigure}
    \caption[Two numerical solutions]{
    A depiction of the inverted pendulum's optimal control profiles of the
    (a) path-homotopy between two different objective functions,
    (b) and policy that switches objective functions parameterised by $\alpha=(0,0.5,1)$ at times $t=(0,2,4)$, respectively.
    Note how  the policy in (b) thus is similar to (a)-solid in the beginning  and (a)-dashed and end, corresponding to a 2-step switch from fuel to time optimal state feedback control learnt by a neural network.
    }
\end{figure}

In recent years, with the progression of machine learning (ML), data-driven approaches have become popular in the context of control. 
Across these approaches, control policies are represented by neural networks (NN) \cite{lillicrap2015continuous}, whose feed-forward architectures express simple executable functions. When given appropriate data, these models can be trained to approximate optimal control policies, and pose a reasonable alternative to MPC in terms of both performance and computational complexity  \cite{aakesson2006neural}, especially on low-end hardware.

Whilst the problem of synthesising optimal control policies has been well addressed in ML literature, in both single and multi-objective contexts \cite{schulman2017proximal, sanchez2018real, liu2014multiobjective}, relatively few have tackled the problem of adapting policies online to changing objectives \cite{natarajan2005dynamic, abels2018dynamic}. Such a case is particularly relevant to systems in which objective priorities can change unexpectedly, e.g. saving energy vs. saving time.
To the best our knowledge, the use of trajectory optimisation and imitation learning has been unexplored in this regard.

In this work, we show that we can straightforwardly leverage trajectory optimisation and homotopy continuation to synthesise state-feedback controllers capable of adapting online to dynamic objectives. Our contributions are:
\begin{enumerate}
    \item we develop an efficient data-generation pipeline to produce optimal demonstrations across multiple objectives and boundary conditions;
    \item and we show that our trained policies can continuously represent homotopies between optimal policies.
\end{enumerate}

The outline of the paper is as follows.
In Section \ref{sec:relatedwork} we describe related work.
Then, in Section \ref{sec:optimalcontrol}, we explain the homotopy continuation method we use for trajectory optimisation.
In Section \ref{sec:learning}, we explain our method of dataset generation, our machine learning setup, and the results we achieved.
Finally, in Section \ref{sec:conclusion}, we reflect upon our results and discuss what they entail.

\section{Related Work}
\label{sec:relatedwork}

Many problems in robotics are naturally formulated in terms of
nonlinear optimal control problems
 \cite{zeng2017energy, izzo2016designing, shen2018trajectory}, but they often lack a closed-form solution.
One can usually, however, solve the trajectory optimisation problem (TOP) using numerical methods \cite{betts2010practical}, to obtain open-loop solutions. But, as described before, na\"ively executing or tracking these solutions might have drawbacks in terms of robustness.

MPC addresses this issue by iteratively computing and executing an open-loop solution on a finite-time horizon, allowing for online reactions in response to unforeseen events.
With advances in MPC's efficiency \cite{wang2010fast} and increasing computational capabilities of many robotic platforms, this approach has become popular in a variety of applications \cite{shen2018trajectory, nair2018robust, baca2018model, neunert2018whole, mayne2014model, zeilinger2014real}.
But, although it can be quite effective, the need to employ a finite-time horizon, regardless of computational capabilities, results in suboptimal trajectories, as they do not take the full-time horizon into account.

A set of approaches that mitigates these issues, by learning the optimal control with respect to the full- (or infinite-) time horizon, uses reinforcement learning (RL) \cite{bertsekas2019reinforcement, sutton2011reinforcement}.
These approaches operate by constructing a policy through incremental interaction with the environment.
Although such approaches have been quite successful in the continuous control domain \cite{lillicrap2015continuous}, particularly for stochastic systems, they sometimes struggle to converge and are sensitive to the choice of rewards \cite{duan2016benchmarking, hadfield2017inverse}.


In the case that expert demonstrations are either readily available or producible, it is sometimes more straightforward to rely on the alternative ML approach: imitation learning (IL).
In this approach, expert demonstrations are utilised to synthesise optimal policies in a supervised-learning manner.
This approach benefits from a greater sampling-efficiency and has been shown to rapidly accelerate policy-convergence in RL approaches \cite{sun2018fast, cheng2018fast}.
It is usually the case that demonstrations are not necessarily optimal, but if, however, the system at hand is well-suited to providing optimal demonstrations, e.g. through trajectory optimisation, this approach proves to be particularly useful in converging to an optimal policy in an uncomplicated way \cite{sanchez2018real, izzo2018machine, tailor2019learning, sprague2018adding}.

A commonality among these approaches --- IL and RL --- is that their policies are represented by NNs and are traditionally orientated towards single-objective scenarios only.
Some pertaining works have indeed addressed the multi-objective\footnote{The single- and multi-objective modalities of trajectory optimisation are often referred to as \textit{single-} and \textit{multi-task} in ML parlance, respectively. Whilst, in ML, these terms often describe boundary conditions associated with different tasks, we maintain that an \textit{objective} describes only the cost of a trajectory ($\mathcal{J}$).}
case, but typically have only sought policies lying along the Pareto front \cite{liu2014multiobjective}, and only few have targeted the case where objective priorities may change over time \cite{natarajan2005dynamic, abels2018dynamic}. 

In the virtual-first work to address this case \cite{natarajan2005dynamic}, a set of policies was learnt in an RL setting, to discretely represent a spectrum of different objectives, which could then be referred to online to adapt to changing objective priorities. 
However, as noted in \cite{abels2018dynamic}, using a discrete set of policies does not scale to complex problems nor truly represent a continuous mapping between objectives.
Recently, \cite{abels2018dynamic} showed that policies could be adapted online to dynamic preferences in an RL setting, using a Q-network, conditioned on the objective-priority.
We diverge from these works in that we show how trajectory optimisation and homotopy can be straightforwardly leveraged to synthesise these conditional policies in an IL setting.

In \cite{sanchez2018real}, it was shown that Pontryagin's minimum principle (PMP)\footnote{PMP is a manifestation of \textit{indirect} trajectory optimisation; alternatively, \textit{direct} methods could be used to generate optimal demonstrations in the case of a black-box dynamics model. For an overview of trajectory optimisation methods refer to \cite{betts2010practical}.} \cite{pontryagin} could be used to generate optimal demonstrations in order to synthesise optimal policies in an IL setting.
It was shown, across a variety of two-dimensional dynamical systems, that the resulting policies, when executed in simulation, produced near-optimal trajectories, even when the systems were initialised in unseen regions of the state-space.
In \cite{izzo2018machine}, we showed that this approach could be extended to three dimensions and used to encode manifold boundary conditions in the state-space, using transversality conditions of PMP.
These results were built upon in \cite{tailor2019learning}, where an investigation of NN hyperparameters and an improved evaluation metric --- \textit{policy trajectory optimality} --- were presented.

In these works, optimal policies were synthesised for a single objective. However, in both \cite{sanchez2018real} and \cite{sprague2018adding}, homotopy continuation \cite{bonnans2008singular} was employed between different objectives to ease the burden of generating demonstrations under objectives which are particularly challenging for gradient-based optimisers.
Such a case arises when the topology of the TOP becomes discontinuous due to the underlying optimal control being \textit{bang-bang}. We build upon these works by taking advantage of the full homotopy path between objectives to straightforwardly synthesise conditional policies, which can react online to changing objective priorities, as we will describe in more detail in the following sections.

\section{Trajectory optimisation} \label{sec:optimalcontrol}




In this section, we outline our optimal-trajectory dataset generation process, used for learning in Section \ref{sec:learning}.

Throughout this paper, we consider nonlinear autonomous dynamical systems of the form $\dot{\pmb{s}} = \pmb{f}(\pmb{s}, \pmb{u})$, with scalar objectives of the form $\mathcal{J} = \int_{t_0}^{t_f} \mathcal{L}(\pmb{u}(t))\dif t$.
With these, one can solve the TOP to determine an optimal trajectory $[\pmb{s}(t), \pmb{u}(t)]^\intercal$ --- an open-loop solution --- given an initial and desired state.
In this work, we formulate this problem as a nonlinear programme and solve it using numerical optimisation \cite{betts2010practical}, where we must determine the optimal \textit{decision vector} $\pmb{z}^\star$ to obtain the optimal trajectory.
For brevity, we denote the TOPs in Sections \ref{sec:pendulum} and \ref{sec:spacecraft} as $P_0$ and $P_1$, respectively.

\subsection{Homotopy continuation}
Under certain objectives, computing an optimal trajectory can be rather difficult; an example is when the optimal control profile $\pmb{u}^\star(t)$ is discontinuous, e.g. in the bang-bang case.
In order to decrease the burden of finding a solution, homotopy continuation, or regularisation techniques, can be employed to achieve convergence. This method is often used to solve singular-control problems \cite{bonnans2008singular}.

To elaborate further: given two different objectives, one can define a linear homotopy between them
\begin{equation}
\label{cost_function}
    \mathcal{J} = (1 - \alpha) \int_{t_0}^{t_f} \mathcal{L}(\pmb{u}(t)) \dif t + \alpha \int_{t_0}^{t_f} \mathcal{L}'(\pmb{u}(t)) \dif t,
\end{equation}
where the \textit{homotopy parameter} $\alpha \in [0,1]$ characterises the objectives' priorities.
Assuming $\mathcal{L}$ characterises the easier objective, we can solve the TOP first with $\alpha = 0$, then use its solution to resolve the same problem with a slightly increased value.
With this process, we iteratively solve the TOP until $\alpha = 1$, at which point we will have converged to a solution under the difficult objective.
Through this process, we obtain a convergent series of solutions --- a \textit{path-homotopy}.
If there are several different objectives, one can extend this approach to multiple homotopies, as we do in Section \ref{sec:spacecraft}.
This process is further described in Algorithm \ref{algo:policyhomotopy}, where the boundary conditions remain constant, $\mathtt{solve}(\pmb{s}_0,\pmb{z}, \alpha$) minimises (\ref{cost_function}) from an initial state $\pmb{s}_0$,
and $\alpha^*$ is the last successful solution parameter.


\begin{algorithm2e}[ht] 
\small
\caption{Criteria homotopy}
\label{algo:policyhomotopy}
\SetKwFunction{FMain}{homotopy\_criteria}
\SetKwProg{Fn}{Function}{:}{}
\Fn{\FMain{$\pmb{s}_0^\star$, $\pmb{z}^\star$, $\alpha^\star$}}{
$\pmb{T}, \alpha \leftarrow \emptyset, \alpha^\star$ \\
\While{$\alpha < 1$}{
    $\pmb{z} = \mathtt{solve}(\pmb{s}_0^\star, \pmb{z}^\star, \alpha)$ \\
    \If{$\mathtt{successful}(\pmb{z})$}{
        $\pmb{z}^\star, \alpha^\star \leftarrow \pmb{z}, \alpha$ \\
        $\pmb{T} \leftarrow \pmb{T} \cup \{(\pmb{s}_0^\star, \pmb{z}^\star, \alpha^\star) \}$ \hfill \tcp{save traj.}
        $\alpha \leftarrow \alpha \in [\alpha^\star, 1]$ \hfill \tcp{increase  $\alpha$} 
    }
    \Else{
        $\alpha \leftarrow \alpha \in [\alpha^\star, \alpha]$ \hfill \tcp{decrease  $\alpha$}
    }}
    \Return $\pmb{T}$
}
\end{algorithm2e}

Similarly, to obtain a database that encompasses a diverse set of states, we can also utilise continuation upon the initial state.
To do this, we slightly perturb the initial state and resolve the TOP with the previous solution as the initial guess, again leading to a convergent series of solutions.
We further describe this process in Algorithm \ref{algo:statehomotopy}, where the homotopy parameter remains constant, $\delta$ defines the perturbation size, and $n$ is the desired number of solutions.
Through these algorithms, we obtain a large set of TOP solutions $\pmb{T}$, that encompass the entire range of homotopy parameters and a broad range of initial states.




\begin{algorithm2e}[ht]
    \small
    \caption{State Homotopy}
    \label{algo:statehomotopy}
    \SetKwFunction{FMain}{homotopy\_state}
    \SetKwProg{Fn}{Function}{:}{}
    \Fn{\FMain{$\pmb{s}_0^\star$, $\pmb{z}^\star$, $\alpha^\star$, $n$}}{
        $\pmb{T} \leftarrow \emptyset$\\
        \While{$|\pmb{T}| < n$}{
            $\pmb{s}_0 \leftarrow \mathtt{perturb}(\pmb{s}_0^\star, ~\delta)$ \\
            $\pmb{z} \leftarrow \mathtt{solve}(\pmb{s}_0,~\pmb{z}^\star, \alpha^\star)$ \\
            \If{$\mathtt{successful}(\pmb{z})$}{
                $\pmb{s}_0^\star, \pmb{z}^\star \leftarrow \pmb{s}_0, \pmb{z}$ \\
                $\pmb{T} \leftarrow \pmb{T} \cup \{(\pmb{s}_0^\star, \pmb{z}^\star, \alpha^\star) \}$ \\
                $\delta \leftarrow \mathtt{increase}(\delta)$
            }
            \Else{
                $\delta \leftarrow \mathtt{decrease}(\delta)$
            }
        }
        \KwRet{\pmb{T}}
    }
\end{algorithm2e}

\subsection{Inverted pendulum swing up} \label{sec:pendulum}
For our first test case, we consider the simple nondimensional inverted pendulum \cite{adding_neural_nets},
\begin{equation} \label{eq:eom}
    \dot{\pmb{s}} = 
    \left[\begin{matrix}\dot{x}\\
    \dot{v}\\
    \dot{\theta}\\
    \dot{\omega}\end{matrix}\right]
    =
    \left[\begin{matrix}
        v\\
        u\\
        \omega\\
        \sin{\left (\theta \right )} - u \cos{\left (\theta \right )}
    \end{matrix}\right] 
    = \pmb{f}(\pmb{s}, u),
\end{equation}
where the variables are: cart position $x$, cart velocity $v$, pole angle $\theta$, pole angular velocity $\omega$, and control $u \in [-1, 1]$.


We begin by defining a homotopic objective
\begin{equation}
\mathcal{J} = \left(1 - \alpha \right)\int_{t_0}^{t_f}u\left(t\right)^{2}\dif t + \alpha \int_{t_0}^{t_f}\dif t ,
\end{equation}
defined from quadratically optimal ($\alpha = 0$) to time-optimal ($\alpha = 1$) control. 
Using PMP, we define the \textit{Hamiltonian}
\begin{multline}
    \mathcal{H} = 
    \pmb{\lambda}^\intercal \cdot \pmb{f}(\pmb{s}, u) + \mathcal{L} = 
    \lambda_{x} v +
    \lambda_{v} u +
    \lambda_{\theta} \omega \\ + 
    \lambda_{\omega} \left(\sin{\left (\theta \right )} - u \cos{\left (\theta \right )} \right) + 
    \left(1 - \alpha \right)u^{2} + \alpha,
\end{multline}
where $\pmb{\lambda}$ is the Lagrange multiplier vector --- or \textit{costate} --- having the same dimensionality as the state.
We then minimise $\mathcal{H}$ with respect to $u$ to find the optimal control
\begin{equation}
    u^\star = 
    \underset{u}{\text{argmin}}\left(\mathcal{H}\right) = 
    \frac{\lambda_{v} - \lambda_{\omega} \cos{\left (\theta \right )}}{2 \left(\alpha - 1\right)},
\end{equation}
defined for $\alpha \in [0, 1)$.
As we drive the homotopy parameter to unity, we find the optimal \textit{switching function}
\begin{equation}
    \sigma =  \lambda_{v} - \lambda_{\omega} \cos{\left (\theta \right )},
\end{equation}
defining the bounded time-optimal control
\begin{equation}
\lim_{\alpha \to 1}\left(u^\star\right) = 
  \begin{cases}
    -1 &  \operatorname{if}~ \sigma < 0 \\
    1 &  \operatorname{if}~ \sigma > 0
  \end{cases},
\end{equation}
where it should be noted that the edge case $\sigma = 0$ is only encountered instantaneously, if at all, so singular control analysis is not necessary here.
Lastly, we compute the \textit{costate equations}
\begin{equation}
\dot{\pmb{\lambda}} = -\nabla_{\pmb{s}} \mathcal{H}
=
\left[\begin{matrix}0\\- \lambda_x\\- \lambda_\omega \left(u \sin{\left (\theta \right )} + \cos{\left (\theta \right )}\right)\\- \lambda_\theta\end{matrix}\right].
\end{equation}
To encode the swingup task into the TOP, we enforce the equality constraint $\pmb{s}(t_f) = \pmb{0}$. 
With the shooting method \cite{betts2010practical}, the decision vector becomes $\pmb{z} = [T, \pmb{\lambda}(t_0)]^\intercal$, where $T = t_f - t_0$ is the trajectory duration and $\pmb{\lambda}(t_0)$ is the initial costate.
Following Algorithm \ref{algo:policyhomotopy} and considering the nominal downright state $\pmb{s}_0 = [0~0~\pi~0]^\intercal$, we compute our initial path-homotopy between quadratically and time-optimal control, as shown in Figure \ref{fig:pendulum_control_homotopy}.


   

\subsection{Spacecraft orbit transfer} \label{sec:spacecraft}

As an increase in complexity of boundary conditions and dimensionality, we now consider the problem of optimising a spacecraft's trajectory from Earth to the orbit of Mars. 
Its heliocentric dynamics are given by
\begin{equation}
    \dot{\pmb{s}} =
    \left[\begin{matrix}
        \dot{\pmb{r}} \\
        \dot{\pmb{v}} \\
        \dot{m} \\
    \end{matrix}\right] =
    \left[\begin{matrix}
    \pmb{v}\\
    \frac{T u}{m} \hat{\pmb{u}} - \frac{\mu}{r^3} \pmb{r} \\
    - \frac{T u}{I_{sp} g_{0}}\end{matrix}\right] =
    \pmb{f}(\pmb{s}, \pmb{u}),
\end{equation}
where the state variables are: position $\pmb{r} = [x, y, z]^\intercal$, velocity $\pmb{v} = [v_x, v_y, v_z]^\intercal$, and mass $m$. 
The controls are the thrust's throttle $u \in [0, 1]$ and direction $\hat{\pmb{u}} = [\hat{u}_x, \hat{u}_y, \hat{u}_z]^\intercal$. The constant parameters governing the dynamics are: maximum thrust capability $T = 0.2 ~[N]$, specific impulse $I_{sp} = 2500 ~[s]$, gravitational acceleration at sea level $g_0 = 9.81 ~[m \cdot s^{-1}]$, and standard gravitational parameter of the sun $\mu = 1.3271\times 10^{20} ~ [m^3 \cdot s^{-2}]$.

We again construct a homotopic objective,
\begin{equation}
    \mathcal{J} = \left(1 - \beta \right) \int_{t_0}^{t_f} u(t)^{2}  \dif t
    +
    \beta \int_{t_0}^{t_f}  \left(\alpha + u(t) \left(1 - \alpha \right)\right)\dif t,
\end{equation}
where we have added a secondary homotopy parameter $\beta \in [0, 1]$ as a means to feasibly arrive to a solution of the problem at $\alpha \in [0, 1], \beta = 1$.
The homotopies between optimal controls are defined as:
quadratic to effort $\alpha = 0, \beta \in [0, 1]$,
quadratic to time $\alpha = 1, \beta \in [0, 1]$,
and effort to time $\alpha \in [0, 1], \beta = 1$.
Using PMP, we then define the Hamiltonian
\begin{multline}
    \mathcal{H} = 
    \pmb{\lambda}_r \cdot \pmb{v} +
    \pmb{\lambda}_v \cdot \left(\frac{T u}{m} \hat{\pmb{u}} - \frac{\mu}{r^3} \pmb{r} \right) + 
    \lambda_m \left(- \frac{T u}{I_{sp} g_{0}} \right)  \\
    + 
    \beta \left(\alpha + u \left(1 - \alpha \right)\right) + u^{2} \left(1 - \beta \right)
\end{multline}
To  find the optimal thrust direction $\hat{\pmb{u}}^\star$, we isolate the portion of $\mathcal{H}$ that depends on it:
$
\frac{T u}{m} \left( \hat{\pmb{u}} \cdot \pmb{\lambda}_v \right)
$.
Considering that $T$, $m$, and $u$ are positive, $\hat{\pmb{u}}^\star$ must be directed opposite of $\pmb{\lambda}_v$ to be a minimiser:
\begin{equation}
    \hat{\pmb{u}}^\star = - \frac{\pmb{\lambda}_v}{\lambda_v}.
\end{equation}
Substituting $\hat{\pmb{u}}^\star$ into $\mathcal{H}$ and minimising it with respect to $u$, we obtain 
\begin{equation}
    u^\star = 
    \frac{1}{2(1-\beta)}
    \left(
    \frac{T \left(\pmb{\lambda}_v \cdot \pmb{\lambda}_v\right)}{ \lambda_v  m }
    +
    \frac{T \lambda_{m} }{ I_{sp}  g_{0} } +
    \beta  \left(\alpha - 1\right)
    \right),
\end{equation}
\begin{figure*}[t]
    \centering
    \begin{subfigure}[b]{0.18\linewidth}
        \includegraphics[width=1\linewidth]{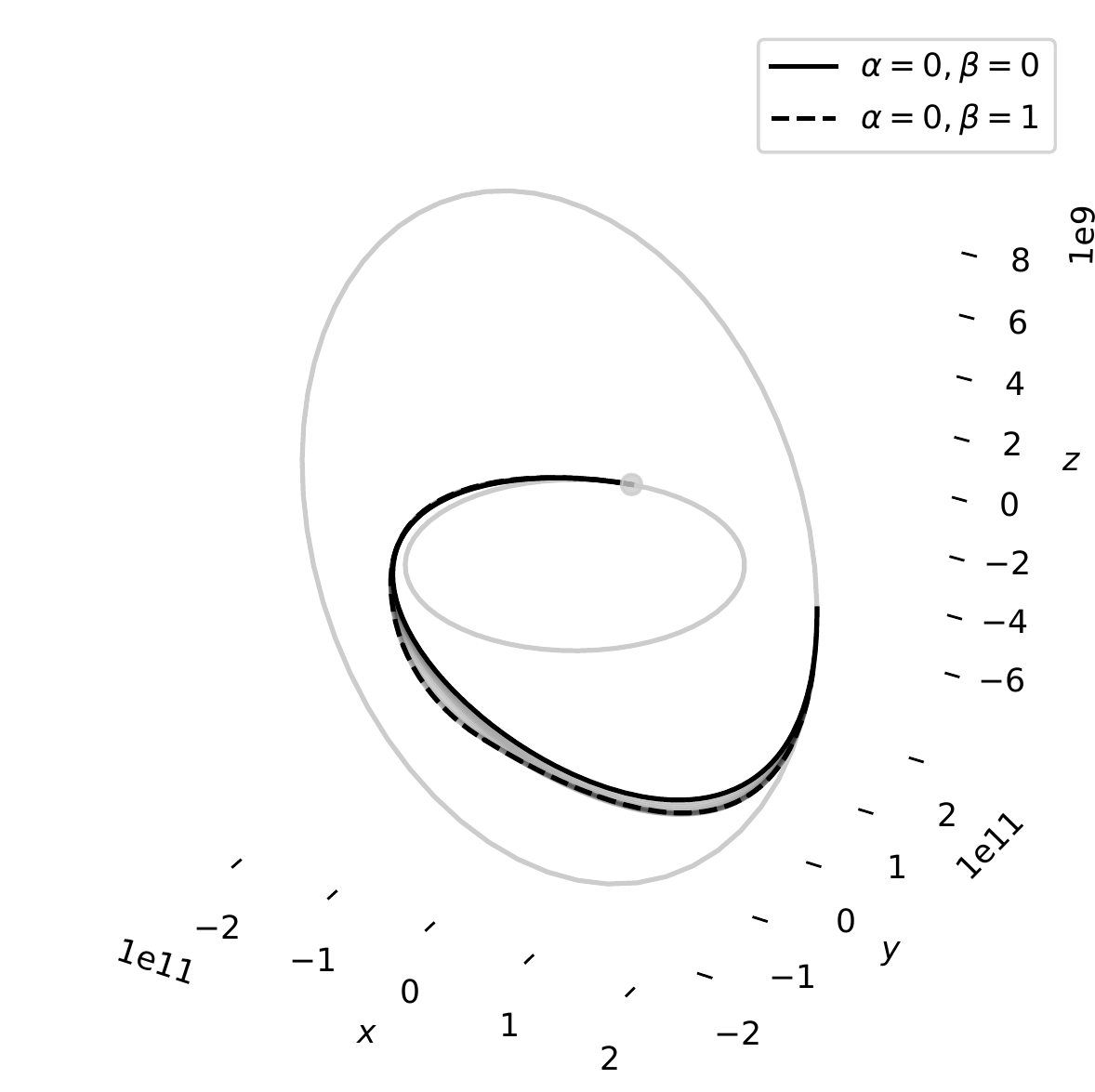}
        \caption{}
        \label{fig:spacecraft_state_homotopy0}
     \end{subfigure}
     \hspace{2mm}
    \begin{subfigure}[b]{0.18\linewidth}
       \includegraphics[width=1\linewidth]{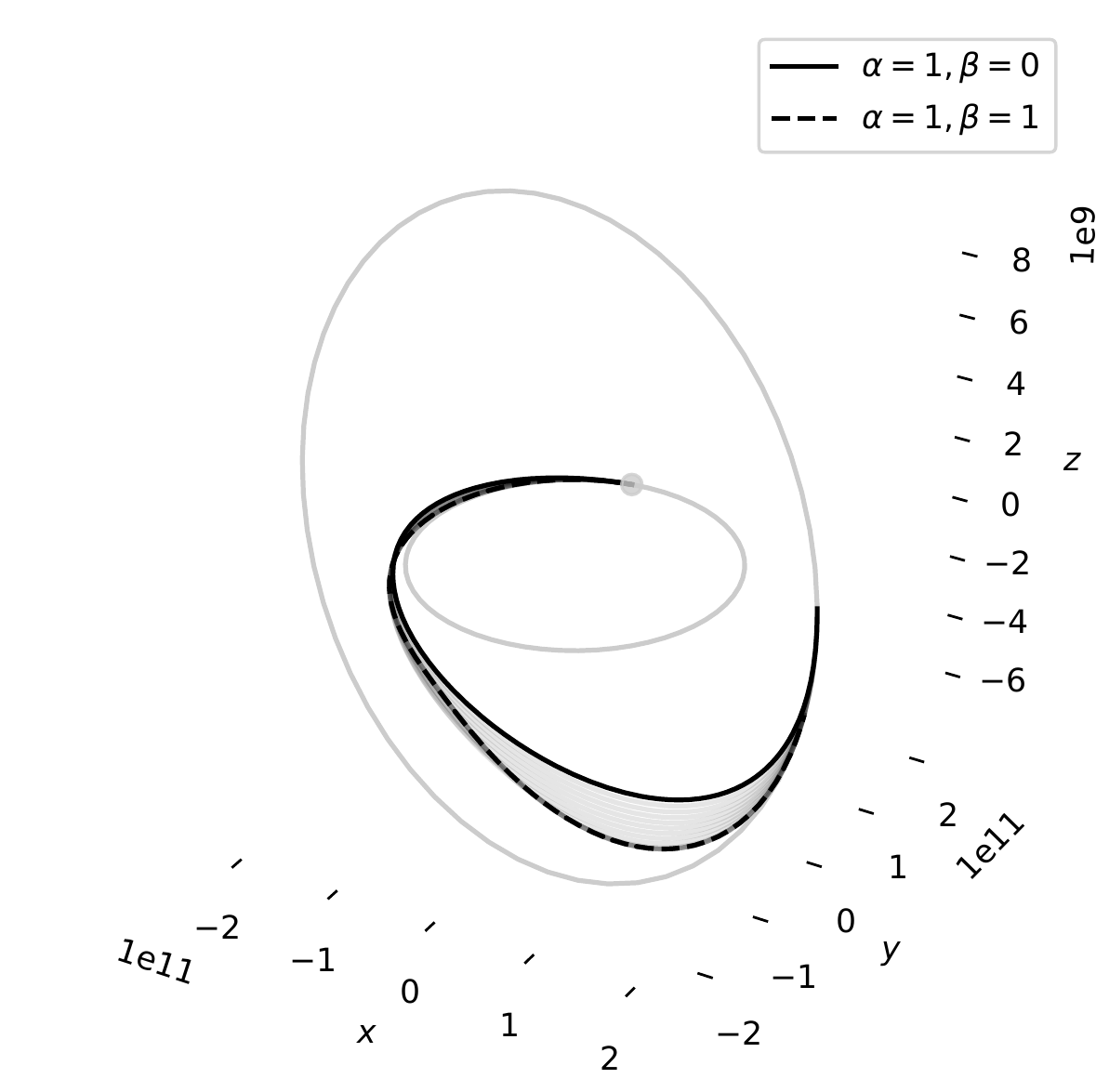}
       \caption{}
       \label{fig:spacecraft_state_homotopy1}
    \end{subfigure}
    \hspace{2mm}
    \begin{subfigure}[b]{0.18\linewidth}
       \includegraphics[width=1\linewidth]{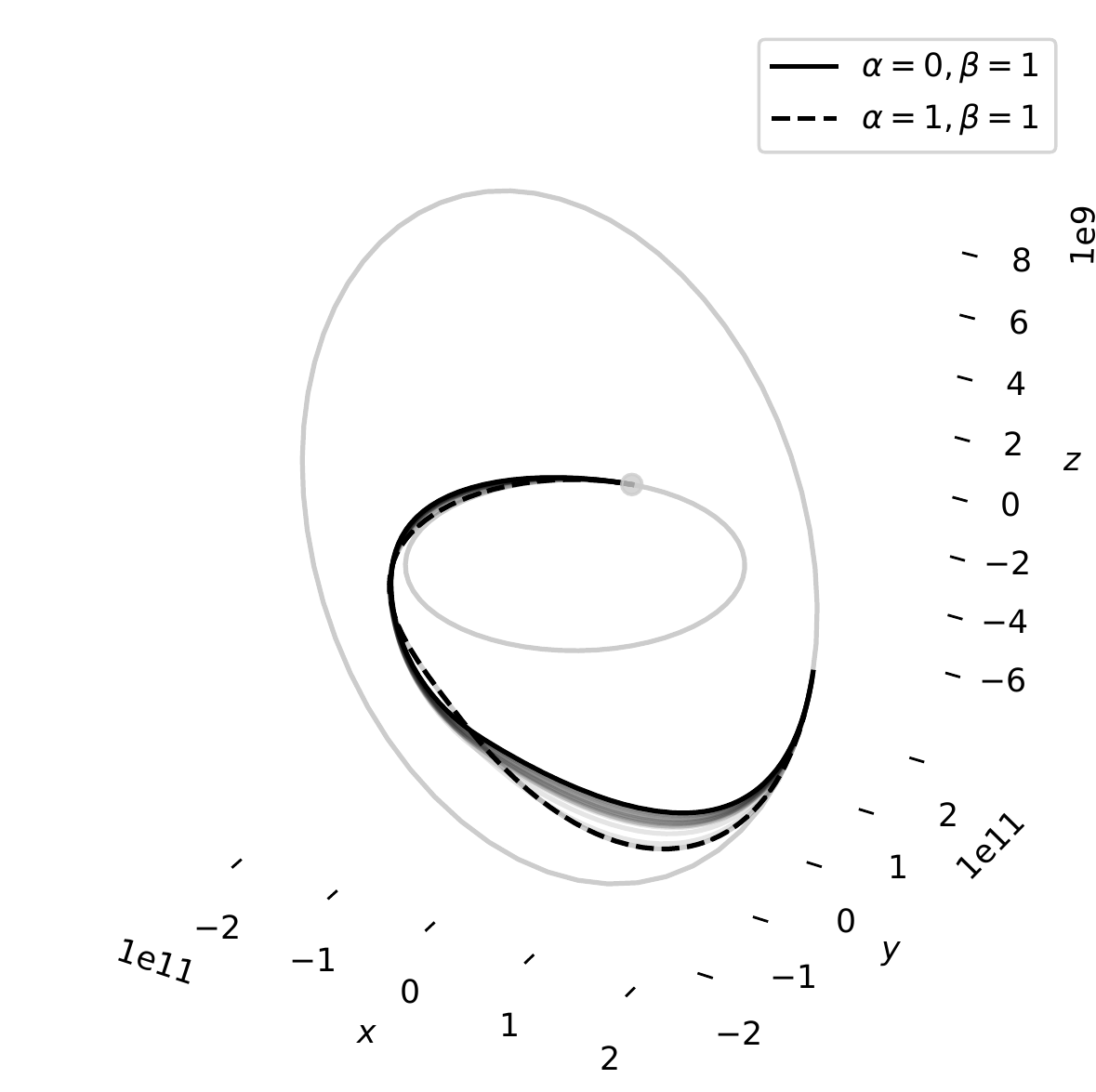}
       \caption{}
       \label{fig:spacecraft_state_homotopy2}
    \end{subfigure}
    \hspace{2mm}
    \begin{subfigure}[b]{0.18\linewidth}
        \includegraphics[width=1\linewidth]{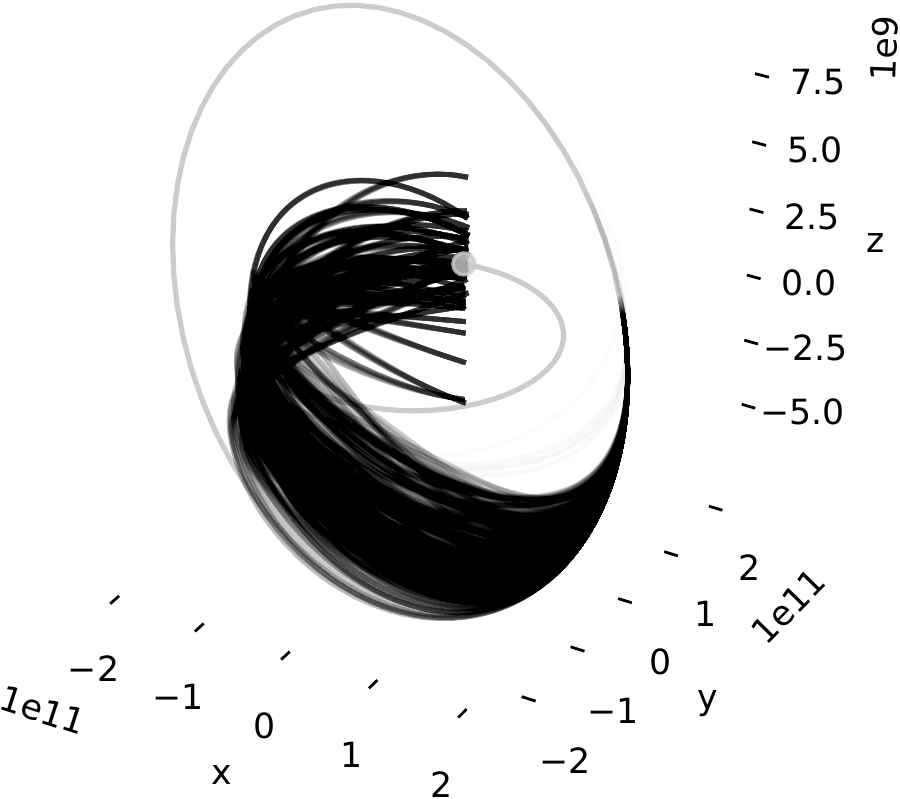}
        \caption{}
    \label{fig:spacecraft_db}
     \end{subfigure}
     \hspace{2mm}
     \begin{subfigure}[b]{0.18\linewidth}
        \includegraphics[width=1\linewidth]{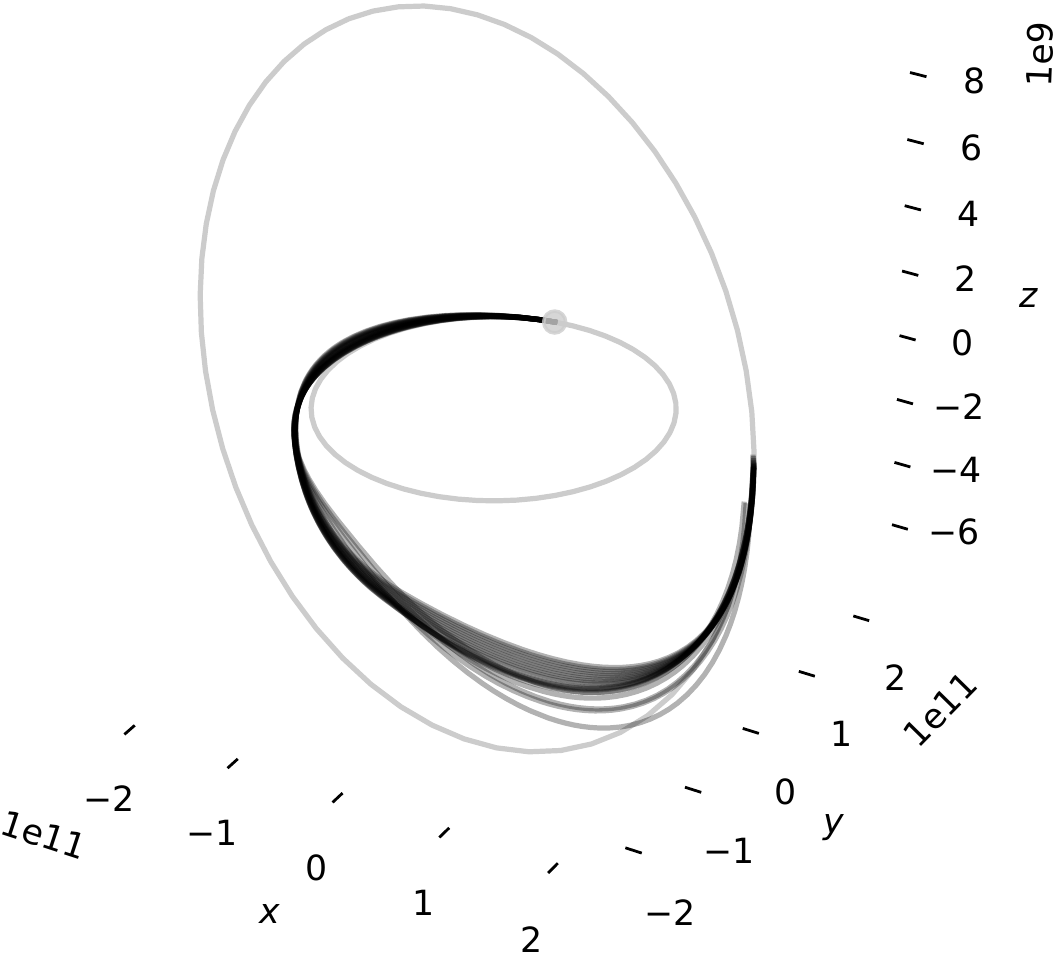}
    \caption{}
    \label{fig:spacecraft_ann_traj}
     \end{subfigure}
    
    \caption[Two numerical solutions]{
        Positional depictions of the spacecraft orbit transfer's ($P_1$)
        path-homotopy between
        (a) quadratically--effort,
        (b) quadratically--time,
        and (c) effort--time optimal control, respectively;
        (d) dataset ($D_1$) of optimal trajectories;
        and (e) policy-controlled trajectories for $\alpha \in [0,1], \beta = 1$.
    }
    \label{fig:spacecraft_state_homotopy}
\end{figure*}
defined for $\alpha \in [0, 1], \beta \in [0, 1)$.
Driving $\beta$ to unity, we find
\begin{equation}
    \sigma = 
    \frac{
        \pmb{\lambda}_v \cdot \pmb{\lambda}_v
    }{
        m \lambda_v
    } +
    \lambda_{m} +  1-\alpha,
\end{equation}
defining the bounded control
\begin{equation}
\lim_{\beta \to 1} \left( u^\star \right) = 
  \begin{cases}
    0 &  \operatorname{if}~ \sigma < 0 \\
    1 &  \operatorname{if}~ \sigma > 0
  \end{cases},
\end{equation}
expressing the homotopy between effort- ($\alpha = 0$) and time-optimal control ($\alpha = 1$), which are discontinuous.
Finally, we compute
\begin{equation}
\dot{\pmb{\lambda}} =
\left[\begin{matrix}
    \frac{\mu}{r^3} \pmb{\lambda}_v - \frac{3\mu}{r^5}\left( \pmb{\lambda}_v \cdot \pmb{r} \right) \\
    - \pmb{\lambda}_r \\
    \frac{T u}{m^2} \left( \pmb{\lambda}_v \cdot \hat{\pmb{u}} \right)
\end{matrix}\right].
\end{equation}
To finish defining the TOP, we enforce two \textit{transversality conditions}.
The first is to allow the optimiser to choose the best final mass
$ \lambda_m(t_f) = 0$,
since it is affected by the control.
The second is to allow the optimiser to choose the best terminal state along Mars's orbit
\begin{equation}
    \frac{
        r^3 \left( \pmb{\lambda}_v \cdot \pmb{v} \right)
        - \mu \left( \pmb{\lambda}_r \cdot \pmb{r} \right)
    }{
        \sqrt{
            \mu^2 \left( \pmb{r} \cdot \pmb{r} \right) +
            \left(\pmb{v} \cdot \pmb{v} \right) \left(\pmb{r} \cdot \pmb{r}\right)^{3}}
    } = 0,
\end{equation}
see \cite{izzo2018machine} for details.
The decision vector becomes 
$\pmb{z} = [T, M(t_f), \pmb{\lambda}(t_0)]^\intercal$, where $T = t_f - t_0$ is the time of flight, $M(t_f)$ is the mean anomaly of the final orbit, and again $\pmb{\lambda}(t_0)$ is the initial costate.
Following Algorithm \ref{algo:policyhomotopy} again and considering Earth's position and velocity as the initial state, and $m(t_0)=1000~[kg]$, we compute each homotopy path, as shown in Figures \ref{fig:spacecraft_state_homotopy0}--\ref{fig:spacecraft_state_homotopy2}, for an interplanetary transfer to somewhere along Mars's orbit\footnote{We pick the starting time to be $0 ~ [MJD2000]$, using Modified Julian Date notation.}.

\section{Learning} \label{sec:learning}
In this section we describe the generated datasets in detail, outline the implemented NN architectures, and evaluate the results of the learning process.

\subsection{Datasets}

To assemble the datasets, we first solve the TOPs from their nominal initial states, as described in Sections \ref{sec:pendulum} and \ref{sec:spacecraft}.
From these initial states, we employ Algorithm \ref{algo:statehomotopy} to obtain a set of solutions $\pmb{T}$ from various other initial states, under the nominal homotopy parameter configurations: $\alpha = 0$ and $\alpha = \beta = 0$ for $P_0$ and $P_1$, respectively.
We then employ Algorithm \ref{algo:policyhomotopy} to obtain solutions across the homotopy parameters' ranges.
Finally, we numerically integrate\footnote{We use an adaptive-stepsize Runge-Kutta method of order 8(5,3) \cite{dormand1980family}.} these systems with the solutions' parameters to obtain datasets of the form $D = \{(\pmb{s}, \alpha, \pmb{u})_i \forall i\}$ --- $D_0$ with $945$ trajectories ($i=577,480$) from $P_0$, and $D_1$ with $5,525$ trajectories ($i=2,124,668$) from $P_1$ ($\beta = 1$).

\subsection{Neural network architectures}
We augment the state-feedback controller formalism to synthesise multilayer-perceptron (MLP) policies of the form $\pi(s, \alpha) \mapsto \pmb{u}^\star$.
We denote $\pi$'s hidden architecture with $m \times n$, describing $n$ hidden layers, each having $m$ nodes.
For every hidden layer we sequentially apply the standard NN operations: linear transformation, layer normalisation, and softplus --- \cite{tailor2019learning} indicated that softplus activations result in smoother and more meaningful control strategies.
For brevity, we denote $P_0$'s and $P_1$'s instances of $\pi$ as $\pi_0$ and $\pi_1$, respectively.
We apply the hyperbolic tangent in the last layer and use its output to approximate the controls of $P_0$ and $P_1$:
$
\pi_0(\pmb{s}, \alpha) \in [-1,1] = u
$
and
$
\pi_1(\pmb{s}, \alpha) \in [-1,1]^3
\mapsto [u~\phi~\theta]^\intercal \in [0,1]\times[0,\pi]\times[0,2\pi]
\mapsto u \cdot [
    \sin(\theta)\cos(\phi),
    \sin(\theta)\sin(\phi),
    \cos(\theta)
]^\intercal
= u \cdot \hat{\pmb{u}}
$,
respectively.

\subsection{Evaluation}
\begin{figure*}[t]
    \begin{subfigure}[t]{0.33\textwidth}
    \centering
       \includegraphics[height=0.64\linewidth]{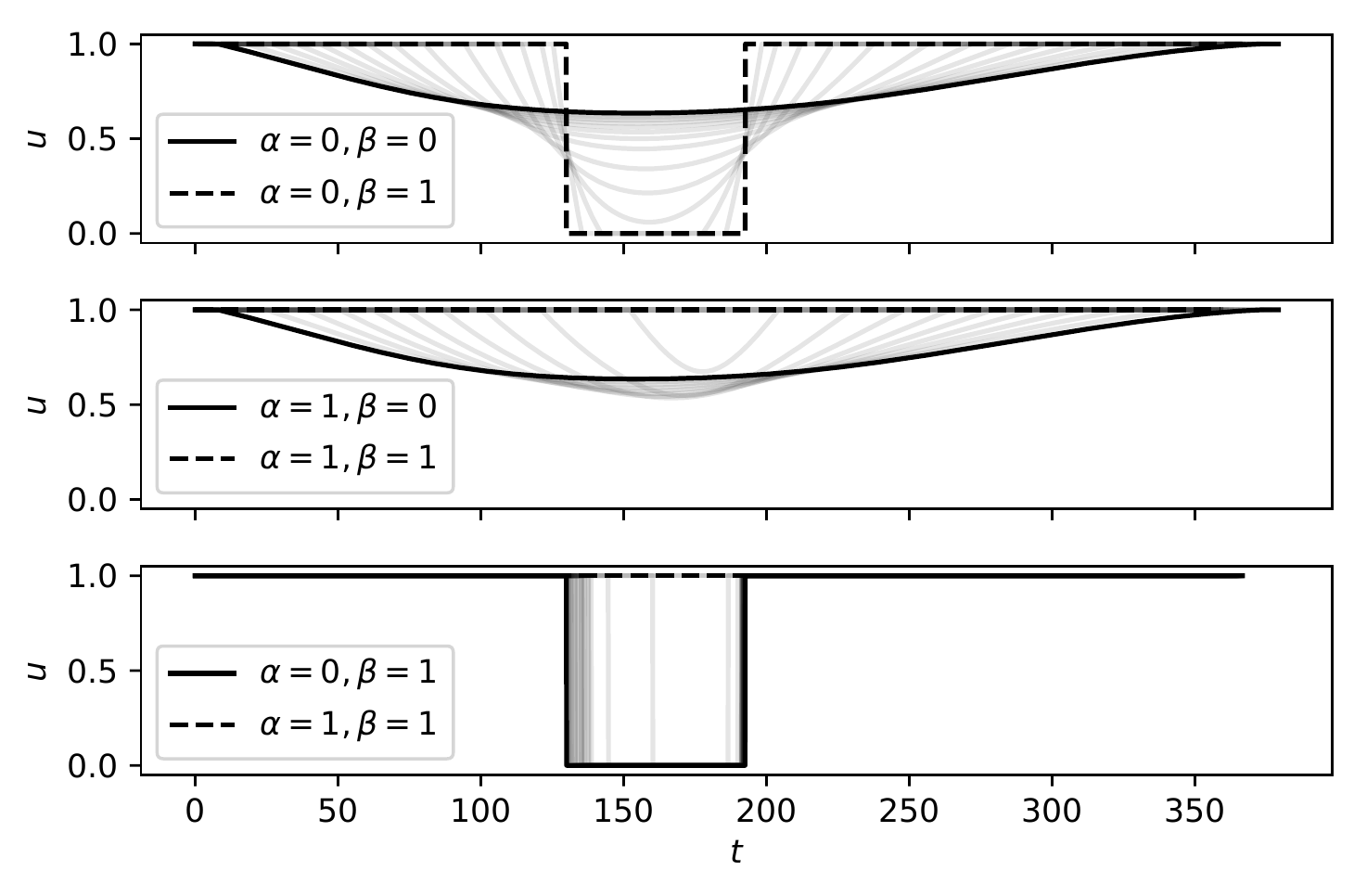}
       \caption{}
       \label{fig:space_nominal}
    \end{subfigure}
    \begin{subfigure}[t]{0.33\textwidth}
    \centering
        \includegraphics[height=0.64\linewidth]{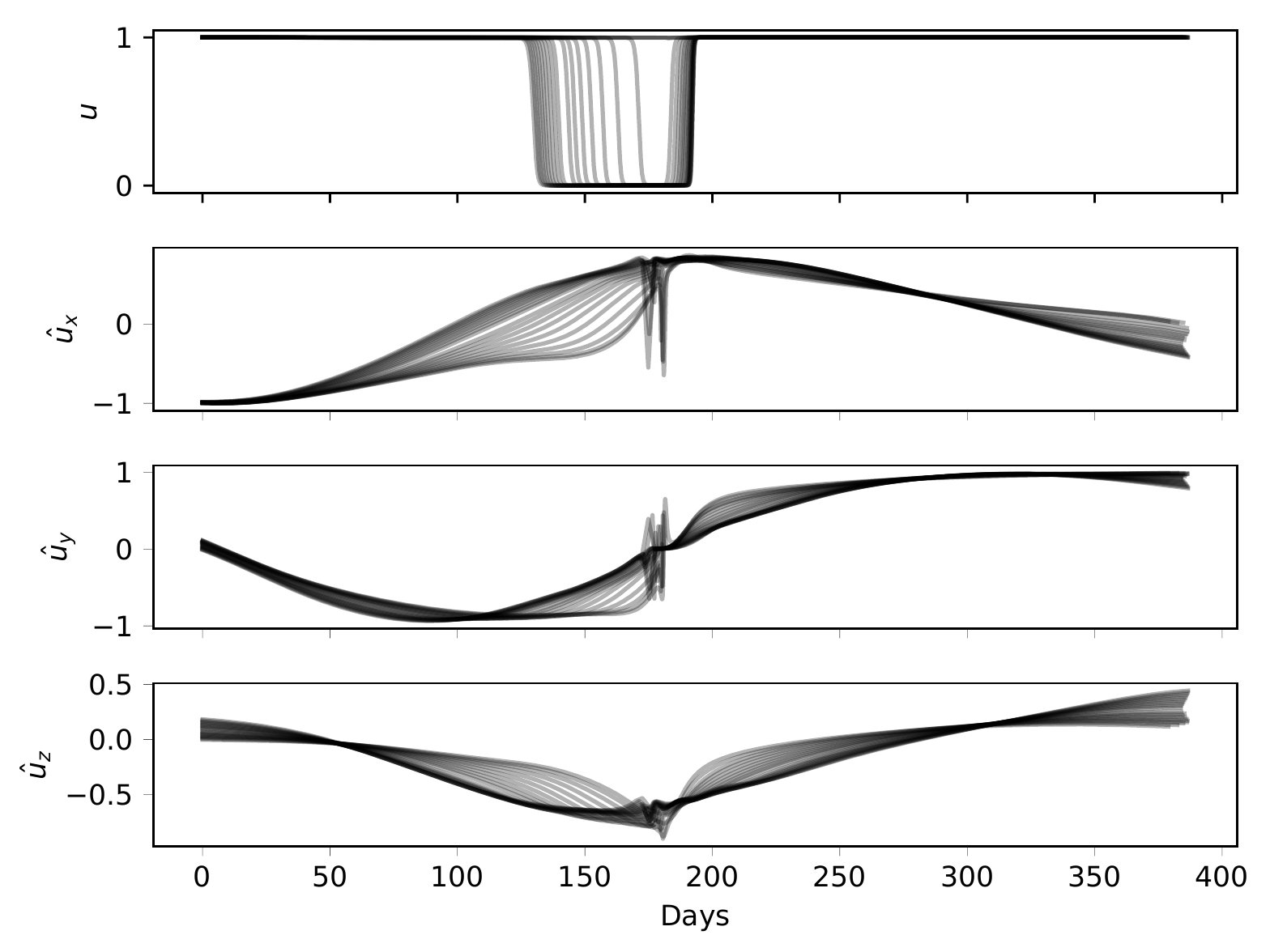}
        \caption{}
        \label{fig:spacecraft_control_homotopy}
     \end{subfigure}
    \begin{subfigure}[t]{0.32\textwidth}
    \centering
       \includegraphics[height=0.64\linewidth]{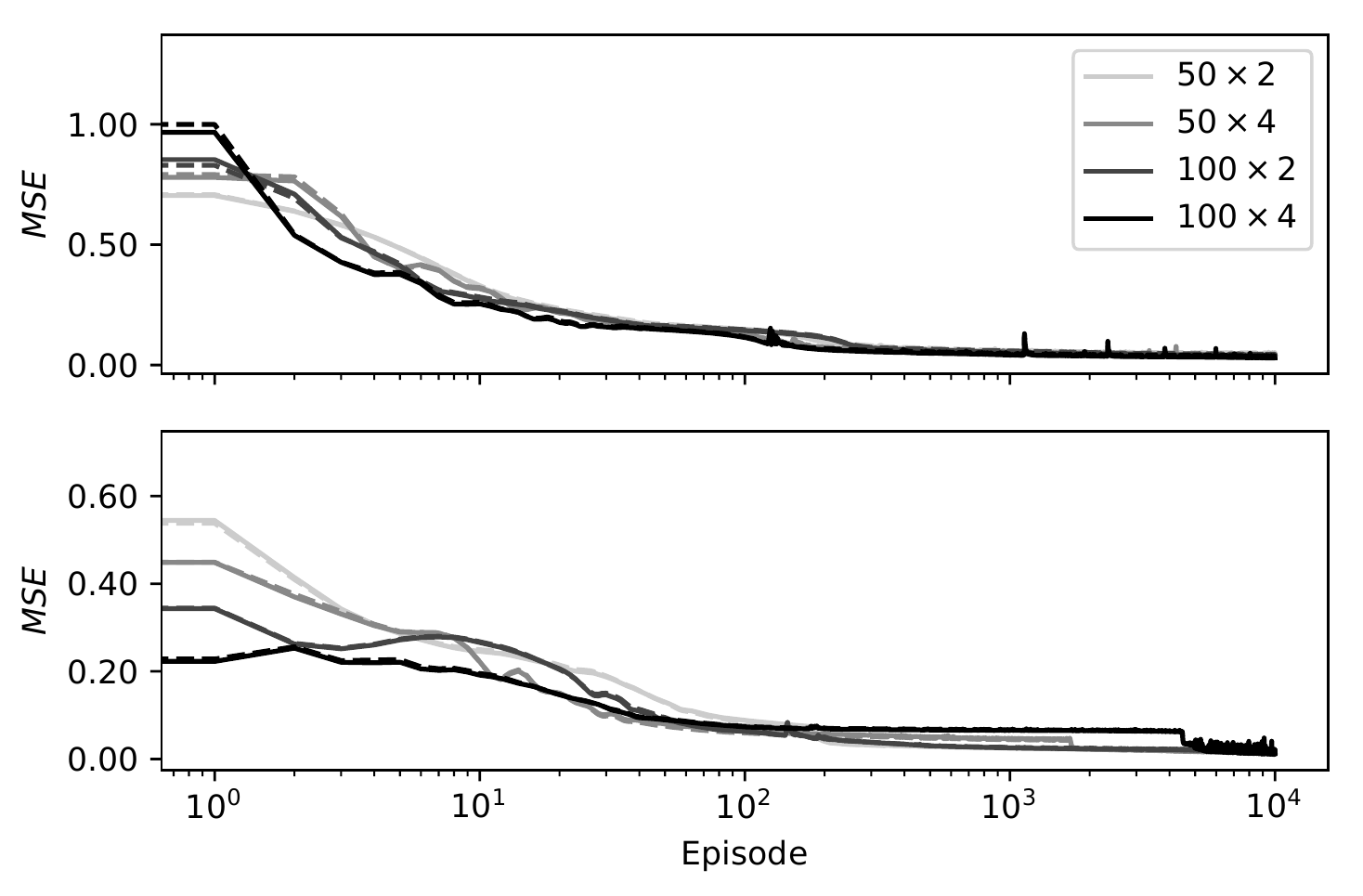}
       \caption{}
       \label{fig:training}
    \end{subfigure}
    \caption{
        (a) Control path-homotopies of $P_1$.
        (b) Control profiles of policy-controlled trajectories in $P_1$ across different homotopy parameters ($\alpha \in [0,1], \beta=1$).
        (c) Training history of $\pi_0$ (top) and $\pi_1$ (bottom).
    }
\end{figure*}

For $\pi_0$ and $\pi_1$, we consider the hidden architectures:
$50 \times 2$, $50 \times 4$, $100 \times 2$, $100 \times 4$.
We train these MLPs on their respective datasets --- $D_0$ and $D_1$ --- for $10,000$ episodes with the Adam optimiser \cite{kingma2014adam}, with a learning rate of $10^{-3}$, a decay rate of $10^{-5}$, and a mean-squared error (MSE) loss function. At each training epoch, we randomly sample $20,000$ datapoints, reserving $10\%$ for validation.
The training history and regression performances are shown in Figure \ref{fig:training} and Table \ref{tab:training}, respectively.

Since success in the regression task does not necessarily mean that the policy-controlled systems will behave as expected, we also evaluate the policy trajectory optimality metric, described in \cite{tailor2019learning}, indicating the optimality-difference between a policy-controlled and optimal trajectory from an initial state.
From the systems' nominal initial states, we evaluate this metric across the hidden architectures and a range of homotopy parameters --- shown in Table \ref{tab:optimality}.




Across all architectures, we find that $\pi_0$ and $\pi_1$ perform well in the regression task and result in trajectories that are near-optimal with respect to the true optimal, across a range of homotopy parameters.
We observe that increasing node breadth and number of layers generally results in slightly better regression performance as well as trajectory optimality.

\begin{table}[]
\begin{tabular}{ccccc}
$\text{Nodes} \times \text{Layers}$ & $\epsilon_0$ & $\epsilon_0'$ & $\epsilon_1$ & $\epsilon_1'$ \\ \hline
$50 \times 2$                       & 0.043434     & 0.049832     & 0.014710     & 0.013711     \\
$50 \times 4$                       & 0.034315     & 0.043264     & 0.014066     & 0.012860     \\
$100 \times 2$                      & 0.038298     & 0.044142     & 0.018373     & 0.018608     \\
$100 \times 4$                      & 0.029807     & 0.036758     & 0.012590     & 0.011232     \\ \hline
\end{tabular}
\caption{The final $MSE$ losses in training ($\epsilon$) and validation ($\epsilon'$) of $\pi_0$ and $\pi_1$ in the regression of $D_0$ and $D_1$, respectively.}
\label{tab:training}
\end{table}

\begin{table}[h]
\centering

\begin{subtable}[h]{\linewidth}
\centering
\begin{tabular}{c|cccc}
{} & \multicolumn{4}{c}{$m \times n$}\\
$\alpha$ &  $50 \times 2$ &  $50 \times 4$ &  $100 \times 2$ &  $100 \times 4$ \\
\hline
0.0 &       8.366217 &       0.201963 &        6.911897 &        0.625839 \\
0.1 &       2.310301 &       2.667611 &        2.722436 &        2.669494 \\
0.2 &       1.270388 &       4.418089 &        1.152010 &        3.045441 \\
0.3 &       3.473601 &       4.555533 &        3.798529 &        5.253961 \\
0.4 &       4.832105 &       3.444665 &        4.429260 &        4.477721 \\
0.5 &       3.260343 &       1.645225 &        2.438281 &        2.761684 \\
0.6 &       1.209908 &       0.616588 &        0.729584 &        1.150946 \\
0.7 &       0.669214 &       0.430044 &        0.279920 &        0.144534 \\
0.8 &       0.497213 &       0.514005 &        0.356340 &        2.509390 \\
0.9 &       0.904339 &       0.797179 &        0.485849 &        0.836367 \\
1.0 &       2.309510 &       1.775131 &        1.027916 &        1.200968 \\
\hline
Mean & 2.645740 & 1.915094 & 2.212002 & 2.243304 \\
\end{tabular}
\caption{Inverted pendulum swingup}
\label{tab:optimality_pendulum}
\end{subtable}
	
\begin{subtable}[h]{\linewidth}
\centering

\begin{tabular}{c|cccc}
{} & \multicolumn{4}{c}{$m \times n$}\\
$\alpha$ &  $50 \times 2$ &  $50 \times 4$ &  $100 \times 2$ &  $100 \times 4$ \\
\hline
0.0 &       5.097919 &       0.163055 &        0.264219 &        0.094422 \\
0.1 &       3.551325 &       0.057574 &        0.109408 &        0.073708 \\
0.2 &       2.160594 &       0.355176 &        0.468579 &        0.353486 \\
0.3 &       0.954027 &       0.665326 &        0.754457 &        0.675646 \\
0.4 &       0.117137 &       0.979295 &        0.975896 &        1.031487 \\
0.5 &       0.519775 &       0.344638 &        0.242022 &        0.455965 \\
0.6 &       0.832056 &       0.210354 &        0.353348 &        0.051248 \\
0.7 &       0.940100 &       0.622716 &        0.765109 &        0.439597 \\
0.8 &       1.231413 &       1.173717 &        1.259935 &        1.005639 \\
0.9 &       0.799833 &       1.096328 &        1.055267 &        1.012649 \\
1.0 &       1.103462 &       0.189937 &        0.768724 &        0.044418 \\
\hline
Mean & 1.573422 & 0.532556 & 0.637906 & 0.476206 \\
\end{tabular}
\caption{Spacecraft orbit transfer}
\label{tab:optimality_spacecraft}
\end{subtable}
	
\caption{
    Policy trajectory optimality gap ($\%$) between the policy-controlled and optimal trajectories for different architectures ($m \times n$) and homotopy parameters ($\alpha$).
}
\label{tab:optimality}
\end{table}

\section{Conclusions} \label{sec:conclusion}
We presented a novel approach to synthesising objective-conditioned near-optimal policies through trajectory optimisation, homotopy, and imitation learning.
We further demonstrated that these state-feedback policies produce trajectories that are near-optimal, and can adapt their behaviours online to changing objectives, due to unexpected events or user-defined preferences.

\section*{Acknowledgement}
This  work  was  supported  by  Stiftelsen  for StrategiskForskning  
(SSF)  through  the  Swedish  Maritime Robotics Centre (SMaRC) (IRC15-0046).

\balance
\bibliographystyle{IEEEtran}
\bibliography{refs}

\end{document}